\documentclass[twocolumn,reprint,superscriptaddress,secnumarabic,showpacs,preprintnumbers,amsmath,amssymb, aps, prd]{revtex4-1}

\usepackage{graphicx}
\usepackage{dcolumn}
\usepackage{bm}
\usepackage{amsmath}
\usepackage{float}
\usepackage{color}
\usepackage{natbib}

\begin{document}

\title{Steering Eco-Evolutionary Games Dynamics with Manifold Control}%


\author{Xin Wang}
\affiliation{LMIB, NLSDE, BDBC, PCL and School of Mathematical Sciences, Beihang University, Beijing 100191, China}
\affiliation{Department of Mathematics, Dartmouth College, Hanover, NH 03755, USA}

\author{Zhiming Zheng}
\affiliation{LMIB, NLSDE, BDBC, PCL and School of Mathematical Sciences, Beihang University, Beijing 100191, China}

\author{Feng Fu}
\email{Feng.Fu@dartmouth.edu}
\affiliation{Department of Mathematics, Dartmouth College, Hanover, NH 03755, USA}
\affiliation{Department of Biomedical Data Science, Geisel School of Medicine at Dartmouth, Lebanon, NH 03756, USA}

\date{\today}%

\begin{abstract}

Feedback loops between population dynamics of individuals and their ecological environment are ubiquitously found in nature, and have shown profound effects on the resulting eco-evolutionary dynamics. Incorporating linear environmental feedback law into replicator dynamics of two-player games, recent theoretical studies shed light on understanding the oscillating dynamics of social dilemma. However, detailed effects of more general \emph{nonlinear} feedback loops in multi-player games, which is more common especially in microbial systems, remain unclear. Here, we focus on ecological public goods games with environmental feedbacks driven by nonlinear selection gradient. Unlike previous models, multiple segments of stable and unstable equilibrium \emph{manifolds} can emerge from the population dynamical systems. We find that a larger relative asymmetrical feedback speed for group interactions centered on cooperators not only accelerates the convergence of stable manifolds, but also increases the attraction basin of these stable manifolds. Furthermore, our work offers an innovative manifold control approach:  by designing appropriate switching control laws, we are able to steer the eco-evolutionary dynamics to any desired population states. Our mathematical framework is an important generalization and complement to  coevolutionary game dynamics, and also fills the theoretical gap in guiding the widespread problem of population state control in microbial experiments. 

%
%
%
%
%
%
 
 \end{abstract}
\keywords{eco-evolutionary dynamics, environmental feedback, switching control, cooperation, social dilemma}

\maketitle


%
%

\section*{Popular Summary}
Changes in environment where individuals interact and compete can drastically impact evolutionary course and outcome in a wide variety of population systems, ranging from microbial cooperation to antibiotic resistance evolution. Such environmental changes are often unprecedented in the nature or simply the result of manual interventions using control devices like chemostat. There has been growing interest in incorporating environmental feedbacks into eco-evolutionary dynamics, yet it remains largely unknown if it is possible (and how) to steer eco-evolutionary dynamics with external switching feedback control laws that adjust selection gradient in the population systems. To fill this theoretical gap, we study eco-evolutionary dynamics of group cooperation with environmental feedbacks that modulate multi-person public goods game interactions. We find the existence of stable equilibrium manifold where the population can settle on and derive potential external control inputs that can steer the population to any desired states. 

In this work, we extend the mathematical framework of eco-evolutionary game dynamics to incorporate realistic asymmetrical environmental feedbacks, for game interactions organized by focal cooperators may have a different efficiency than the ones by defectors. Because of such complex interactions, multiple segments of stable and unstable manifolds can emerge from the population dynamical systems. Our work is in line with previous experimental work demonstrating the existence of (unstable) manifold (`separatrix') in population systems. Our results further demonstrate that the stability of these manifolds can be manipulated by designing population-state dependent switching control laws that tune the nonlinear selection gradient in order to steer the population system to enter any desired states.

Our work combines control theory with evolutionary game dynamics, and provides deep insight into methods for (i) stabilizing equilibrium manifold emerging from group cooperation, and more importantly, (ii) conceiving switching control laws that can steer the system to reach any desired states. Although our present study is focused on group cooperation in multi-person public goods games, our results on manifold control are applicable to many other important situations, such as balancing excitatory and inhibitory interactions in neuronal populations and suppressing evolution of drug resistance in cancer treatment, just to name a few.

\section{Introduction}
The feedback between environment and evolutionary dynamics is widespread in a large number of natural systems~\cite{frank1998foundations,hauert2006evolutionary,west2006social,wakano2009spatial,gore2009snowdrift}. In human society, a depleted environment or resource state favors cooperation which results in the mutual growth for both environmental state and cooperators, while subsequently the free-riders increase which in turn leads to the the degradation of the environment~\cite{stewart2014collapse,cortez2018destabilizing}. Understanding this oscillating system dynamics is of vital importance when dealing with the world-wide problems in human society, ranging from overgrazing of common pasture land, overfishing, to some big challenges like pollution control and global warming~\cite{hardin1968tragedy,pauly2002towards,kraak2011exploring,cohen1995population,hauser2014cooperating}. Similar joint effects can also be obtained in some psychological-economic systems, such as social welfare, overuse of antibiotics and anti-vaccine problems~\cite{fu2018social,bauch2003group,bauch2004vaccination,chen2019imperfect}. 

In particular, similar eco-evolutionary feedback loops exist broadly in microbial systems, which has aroused great concern in evolutionary biology and systems biology in recent years~\cite{schoener2011newest,hauert2008ecological,post2009eco,hanski2011eco}.  Among microbes, the cooperation often emerges due to the secretion or the release of public goods, such as extracellular enzymes or extracellular antibiotic compounds~\cite{riley2002bacteriocins,west2003cooperation,kummerli2010molecular,nadell2008sociobiology,levin2014public}.  A fundamental problem is how these bidirectional feedbacks between ecology and evolutionary dynamics affect the emergence of long-term existence of cooperation as well as the corresponding ecological consequences, which is of particular interest in both biology and ecology~\cite{raymond2012dynamics,wintermute2010emergent,hanski2011eco,becks2012functional,turcotte2011impact,cremer2011evolutionary}. An experimental study confirms the existence of strong feedback loop between laboratory yeast population dynamics and the evolutionary dynamics of the SUC2 gene which can mediate the cooperative growth of budding yeast~\cite{sanchez2013feedback}. This feedback can even determine the demographic fate of those social microbial populations. Besides population density, there are many other ecological properties that have been found to play an important role in such reciprocal feedbacks, such as spatial structures of the population, resource regeneration and supply capacity~\cite{wakano2009spatial,gore2009snowdrift,nowak2006five,brockhurst2008resource,ross2009density}. 
 
Furthermore, to give a clear sight into these feedback-evolving games, a theoretical framework called coevolutionary game theory is proposed to analyze the coupled evolution of strategies and environment~\cite{hilbe2018evolution,szolnoki2018environmental,akcay2018collapse,estrela2018environmentally,shao2019evolutionary}. The core idea is to incorporate the game-environment feedback mechanism into replicator dynamics, in which the feedback changes the payoff structure and further influences the evolution of strategies~\cite{weitz2016oscillating,tilman2019evolutionary}. Such framework successfully shows the emergence of an oscillating tragedy of the commons. Similar cycles are also confirmed in asymmetric evolutionary games with heterogenous environment~\cite{hauert2019asymmetric}. As a meaningful example of application, the framework is used for exploring the effects of intrinsic growing capacity of the resources with punishment and inspection mechanisms~\cite{chen2018punishment}. In summary, these eco-evolutionary models reveal the great role the feedback loop plays in resolving social dilemma and promoting the emergence of long-term cooperation.  

However, it has been proved that in most microbial systems, the essential factor that creates density-dependent (or other ecological property-dependent) selection which leads to the existence of feedback loop is the preferential access to the common good for cooperators~\cite{celiker2012competition,koschwanez2011sucrose}. Such preferential access mechanism related to selection direction indicates an important fact that the multiplication factor of cooperators in public goods game (PGG) may actually be influenced by how well they fare against defectors (namely, natural selection gradient in the population). This process in turn affects the evolutionary dynamics, and leads to the existence of an asymmetrical feedback. While most of the previous works focus on two-player games with environment feedback~\cite{weitz2016oscillating,hauert2019asymmetric}, this selection gradient engineering via asymmetrical feedback in PGG, which is more general in microbial systems, is ignored; the effects for this feedback loop remains unclear. In addition, there still lacks a proper understanding of nonlinear feedback mechanisms since most studies concentrate on linear feedback laws. Further, when looking into the current experimental results in microbial systems, we find that there exists a big theoretical gap on how to effectively steer a given initial population state to a desired final state in coevolutionary games dynamics, which may have wide applications in systems biology.  

To advance all these important issues, in this work, we propose a general framework which extends the two-player games with environmental feedback to coevolutionary multi-player games with asymmetrical feedback driven by nonlinear selection gradient. Inspired by the experimental results in ~\cite{sanchez2013feedback} that reveals the existence of strong feedback loop between population strategy and population density as well as the emergence of `separatrix' line in dynamics, we shall provide a general model to describe similar eco-evolutionary dynamics by considering ecological properties solely being affected by the fraction of cooperators as a benefit of cooperation, which can be described by the multiplication factor of cooperators. Unusually, we find the emergence of multiple segments of stable and unstable manifolds with a number of different feedback control functions, which is a new phenomenon that is totally different from the solely interior equilibrium situations obtained in previous models. It is also worthy of noting that the equilibrium curve of unstable manifold circumstance in our model is in line with the separatrix obtained by experimental results in Ref.~\cite{sanchez2013feedback}, which indicate the potential power for our general framework to explain and understand the eco-evolutionary dynamics in a large amount of the real microbial systems. Furthermore, a larger relative changing speed of the asymmetrical feedback can not only accelerate the convergence speed of stable manifolds, but also increase the attraction basin of the stable manifolds. Our model reveals the detailed effects of nonlinear selection gradient engineering feedback loop in PGG, which is an important complement and generalization for the previous coevolutionary framework. 

Moreover, we highlight the conclusion that when incorporating time-dependent or state-dependent switching laws which can actually controls the stability of the possible manifolds in our framework~\cite{liberzon2003switching,paarporn2018optimal,garone2010switching}, i.e., using manifold control, we can steer the coevolutionary games dynamics to any desired region. Therefore, our framework can be widely applied into culture refresh modes for establishing continuous culture devices and designing needed chemostat for microbial experiments, which is of great significance in systems biology and microbial ecology~\cite{chuang2010decade,winder2011use,matteau2015small}.   

\section{Modeling Framework}
\begin{figure*}[htbp]
    \centering
    \includegraphics[width=0.95\linewidth]{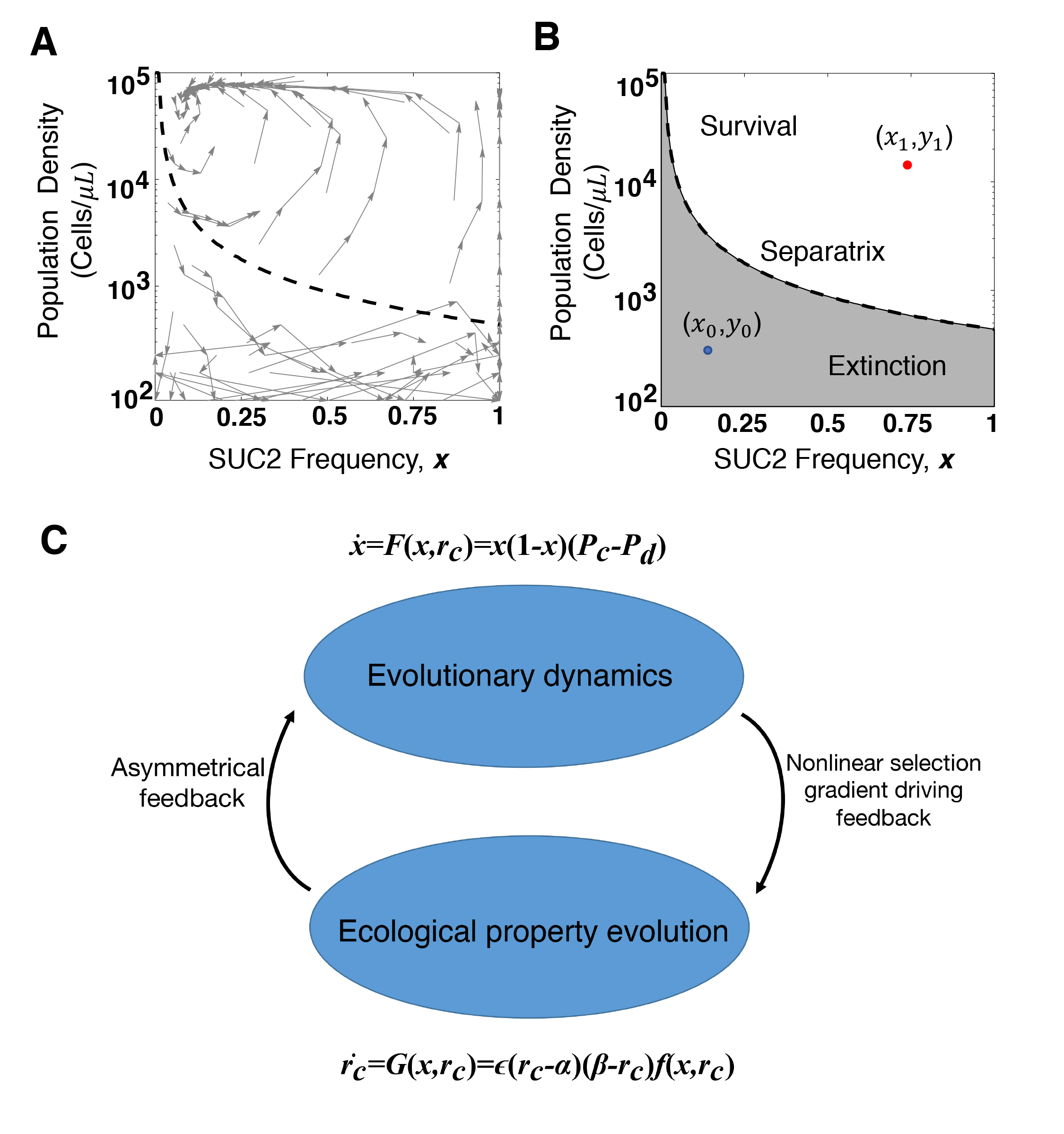}
    \caption{Experiment results in ~\cite{sanchez2013feedback} and schematic of our eco-evolutionary model framework. (A) Eco-evolutionary trajectories of yeast population density and the fraction of cooperators, SUC2 gene. This experimental result consists of 60 cultures over five growth-dilution cycles. The phase graph shows clearly two regions divided by a separatrix line, over which the system converges to an eco-evolutionary equilibrium, while under which the population goes extinct. The separatrix line is predicted by a bi-phasic logistic model, provided in ~\cite{sanchez2013feedback}. (B) Theoretical gap on control problems of coevolutionary dynamics in microbial systems: how to steer the given initial population state $(x_0, y_0)$ to a desired final state $(x_1, y_1)$ with external feedback control laws. (C) Schematic of model framework: eco-evolutionary games with asymmetrical feedback driven by nonlinear selection gradient in public goods game.}
    \label{exp}
\end{figure*}

Here we propose a general framework for evolutionary dynamics with feedback loops in PGG. Our model is inspired by experimental results in ~\cite{sanchez2013feedback} which confirms the existence of strong feedback loop between laboratory yeast population dynamics and the evolutionary dynamics of cooperators, SUC2 gene, as shown in Fig. \ref{exp}(A). Similar feedback loops originating from the preferential access to the public goods for cooperators has also been proved to exist in many microbial systems. The phase graph of the coevolutionary system is clearly divided into two regions by the separatrix line, over which the system converges to an eco-evolutionary equilibrium and the population survives, while under which the population goes extinct. Of particular interest, a specific bi-phasic logistic mathematical model has been proposed to reproduce the emergence of separatrix line, in which the population growth rates of cooperators and defectors are distinguished under different cooperator densities. However, we still lack of a general framework in multi-player games for describing similar coevolutionary dynamics of population strategy and different ecological properties, in a way that population density in Fig. \ref{exp}(A) can be included as an example. Meanwhile, current coevolutionary game theory frameworks exclusively focus on linear feedback laws, the detailed effects of nonlinearity in feedback loops remain unclear. Moreover, in Fig. \ref{exp}(B), we raise an important control problem in general coevolutionary games dynamics which has wide applications in similar microbial experiments: how to effectively steer the given initial population state $(x_0, y_0)$ to a desired final state $(x_1, y_1)$ with external feedback control laws? To make progress on all these unsolved issues, in what follows we present a novel coevolutionary model with feedback laws that can be engineered based on nonlinear selection gradient in the system. 

Consider a well-mixed population. An individual finds itself in a group of size $S+1$, with $S$ other players participating in the PGG. All players can choose to be either a cooperator who contributes $c$ to the public pool or a defector who free rides others' efforts in the group. In classical PGG, the total contributions are multiplied by a multiplication factor $r$ and then divided equally among all participants. In this work, to describe the phenomenon that cooperators have preferential access to the common good in real microbial systems which facilitates the formation of an asymmetrical feedback loop, we assume the multiplication factors of cooperators and defectors are different, denoted by $r_c$ and $r_d$ respectively, and $r_c\ge r_d$. Without loss of generality, we keep $r_d$ constant and let $r_c$ change based on nonlinear selection gradient feedback~\cite{conner2004primer}. In turn, $r_c$ affects the relationship of cooperator's and defector's payoffs and further drives evolutionary dynamics. In this way, we characterize all kinds of ecological properties affected by the fraction of cooperators as a benefit of cooperation that can be described by cooperator's multiplication factor, which makes our framework more general. The schematic of this coevolutionary game framework is shown in Fig. \ref{exp}(C).

Assume $x$ denotes the frequency of cooperators in the population. For any given focal individual, the chance that $k$ out of other $S$ individuals are cooperators is 
\begin{equation*}
	\binom{S}{k}x^k(1-x)^{S-k}.
\end{equation*}
For simplicity and without loss of generality, we set cooperator's cost $c$ equal to $1$. Therefore the expected payoffs of cooperators and defectors, $P_c$ and $P_d$, are
\begin{equation}
    \begin{split}
        P_c &= \sum\limits_{k=0}^{S}\binom{S}{k}x^k(1-x)^{S-k}\left[\frac{(k+1)r_c}{S+1}-1\right] \\
        &= \frac{1+Sx}{S+1}r_c -1 \\
        P_d &= \sum\limits_{k=0}^{S}\binom{S}{k}x^k(1-x)^{S-k}\frac{kr_d}{S+1}\\
        &= \frac{Sx}{S+1}r_d
        \label{e2}
    \end{split}
\end{equation}
Then the replicator dynamics for the fraction of cooperators $x$ is
\begin{equation}
	\begin{split}
		\dot{x} &= x(1-x)(P_c(x, r_c)-P_d(x))\\
		&= x(1-x)\left(\frac{Sx+1}{S+1}r_c-1-\frac{Sx}{S+1}r_d\right)
	\end{split}
\end{equation}
Meanwhile, the feedback-evolving dynamics, i.e., ecological property evolution in Fig. \ref{exp}(C) is given by
\begin{equation}
	\dot{r_c} = \epsilon (r_c-\alpha)(\beta-r_c) f(x, r_c)
\end{equation}
where $\epsilon \ge0$ denotes the relative changing speed of \(r_c\), the cooperator's multiplication factor, compared to strategy dynamics. The logistic term \((r_c-\alpha)(\beta-r_c)\) ensures that $r_c$ is restrained to the range $[\alpha, \beta]$, which satisfies \(1<\alpha<\beta<S+1\) according to the social dilemma in PGG. In addition, $f(x, r_c)$ is a control function that describes the asymmetrical feedback mechanisms in our model. While $f$ actually characterizes the current impact of population strategies on environment, previous works exclusively focus on linear selection gradient feedback laws, such as $f=\theta x-(1-x)$ in ~\cite{weitz2016oscillating} and $f=e_L x+e_H (1-x)$ in ~\cite{tilman2019evolutionary}. Here, we stress our effects on generality of nonlinearity in feedback control laws, the general form of which is: 
\begin{equation}
\begin{split}
	f(x, r_c) = &\Phi_0(x, r_c)(\Phi_1(x, r_c)-a_1)*\\
	      & (\Phi_2(x, r_c)-a_2)...(a_n-\Phi_n(x, r_c)),
 \end{split}
\end{equation}
in which
\begin{equation}
	\begin{split}
		\Phi_0(x, r_c) &= P_c-P_d\\
		\Phi_i(x, r_c) &= \theta_i P_c-P_d
	\end{split}
\end{equation}
and $a_i\ge 0$, $\theta_i>0$, $n+1$ denotes the order of the control function. In particular, $\Phi_0(x, r_c)= P_c-P_d$ is a natural and more general form of linear selection gradient.

Finally, our generalized framework of multi-player evolutionary games with asymmetrical feedback driven by nonlinear selection gradient can be written as follows: 
\begin{equation}
	\begin{cases}
	\begin{split}
		\dot{x} = &x(1-x)\left(P_c-P_d\right)\\
		\dot{r_c} = &\epsilon (r_c-\alpha)(\beta-r_c)(P_c-P_d)*\\
		           & ((\theta_1 P_c-P_d)-a_1)...(a_n-(\theta_n P_c-P_d))
	\end{split}
	\end{cases}
\end{equation}

\section{Results}

\subsection{Emergence of multiple segments of stable and unstable equilibrium manifolds}

Firstly, we analyze the simplest circumstance in which the feedback control term $f$ is a quadratic function, i.e., $n=1$:
\begin{equation}
	f(x, r_c) = (P_c-P_d)(a_1-(\theta_1 P_c-P_d))
\end{equation}
The co-evolutionary model is explicitly described by:
\begin{equation}
	\begin{cases}
	\begin{split}
		\dot{x} &= x(1-x)\left(P_c-P_d\right)\\
		\dot{r_c} &= \epsilon (r_c-\alpha)(\beta-r_c)(P_c-P_d)(a_1-(\theta_1 P_c-P_d))
	\end{split}
	\end{cases}
	\label{e1}
\end{equation}

\begin{figure*}[htbp]
    \centering
    \includegraphics[width=0.95\linewidth]{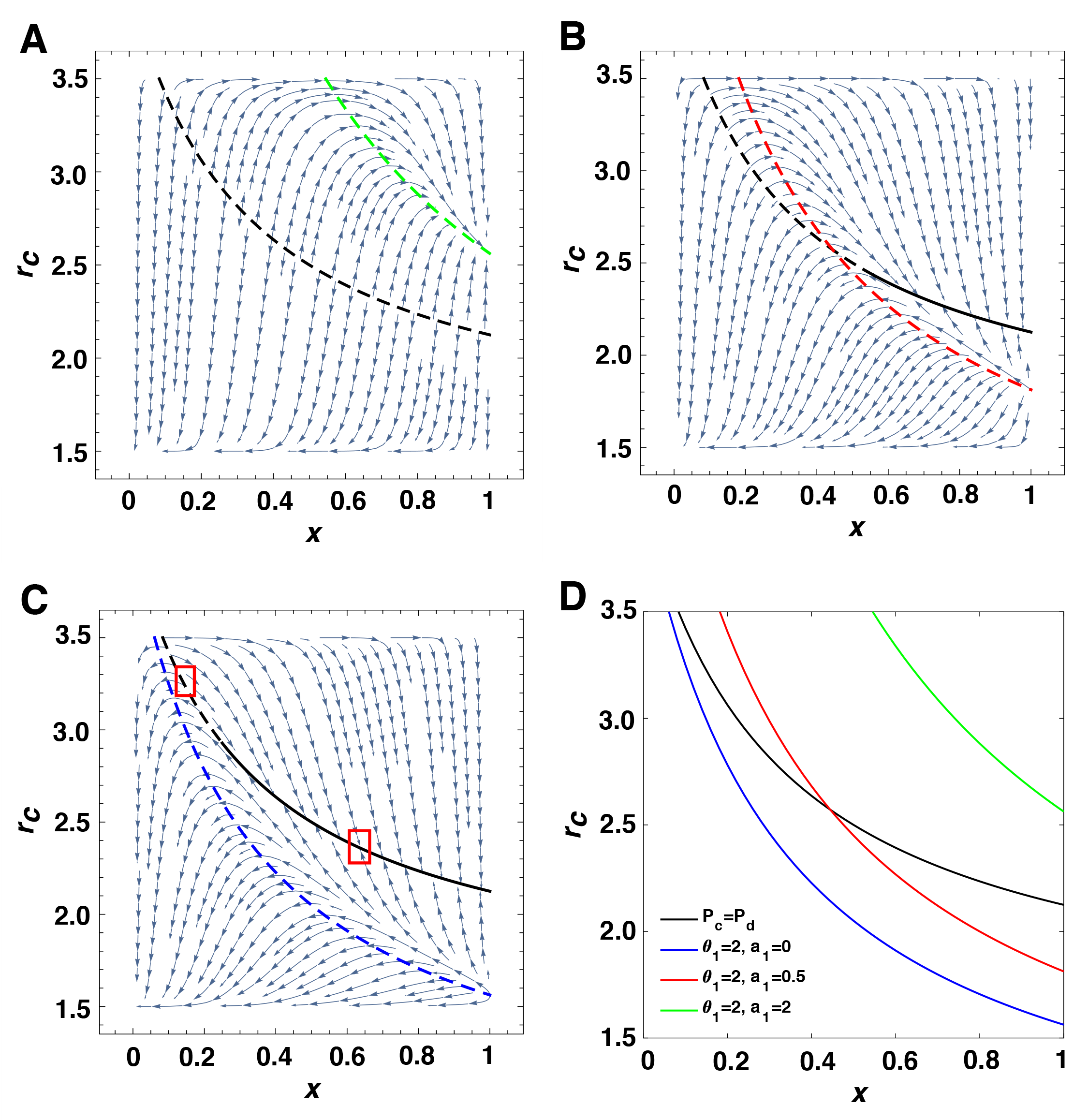}
    \caption{Emergence of stable equilibrium manifolds when feedback control function $f$ is in quadratic forms. (A)-(C) show phase graph under different control conditions. In all subfigures, $\alpha=1.5$, $\beta=3.5$, $S=3$,  $r_d=1.5$, $\epsilon=2$, $\theta_1=2$. We change $a_1=2, 0.5, 0$ respectively. The stable and unstable part of equilibrium curve are indicated by solid and dashed lines, in separate. In particular, in (C), we mark the typical stable and unstable manifold situations in a small neighborhood of the equilibrium curve with red outlines. (D) presents all curves that satisfy $f=0$ in (A)-(C).}
    \label{2}
\end{figure*}

\begin{figure}[htbp]
    \centering
    \includegraphics[width=0.95\linewidth]{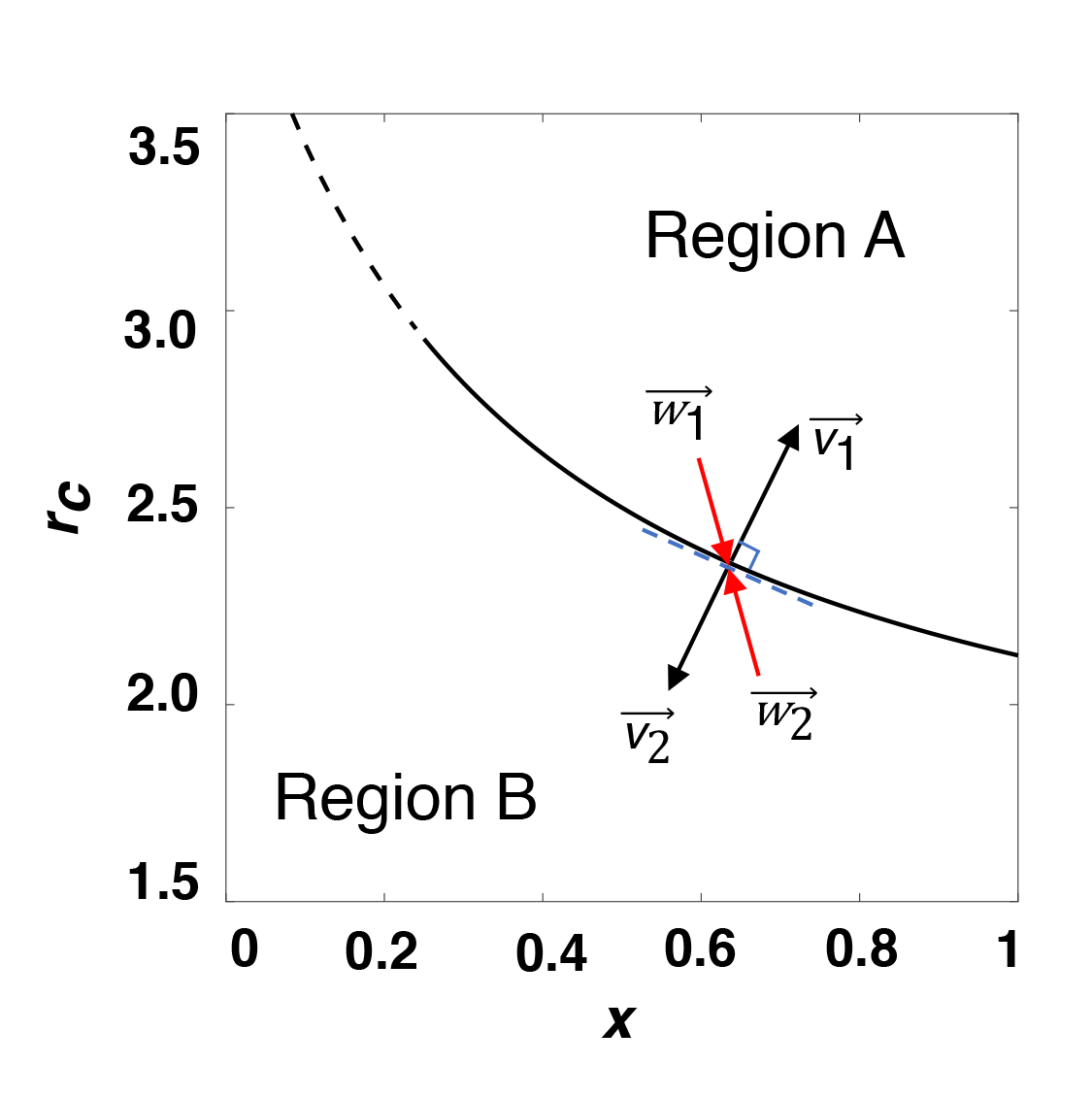}
    \caption{Schematic of stability proof. The fixed point on the equilibrium curve is stable if and only if $\vec{w_i} \cdot \vec{v_i} < 0$ in a small neighborhood, where $\vec{v_i}$ is the normal vector of the equilibrium curve while $\vec{w_i}$ is the trajectory field direction.}
    \label{proof}
\end{figure}

In Fig. \ref{2}, we show the emergence of multiple segments of stable and unstable equilibrium manifolds using phase graphs under different conditions. The parameters are as follows: $\alpha=1.5, \beta=3.5, S=3, r_d=1.5, \epsilon=2, \theta_1=2$ and $a_1=2, 0.5, 0$ respectively. In addition, in Fig. \ref{2}(D), we present all curves that satisfy $f=0$ in Fig. \ref{2}(A)-(C). In the following parts of this paper, to clearly describe the stability of fixed points, we name $P_c=P_d$ as the equilibrium curve and $\theta_i P_c-P_d=a_i$ the control curves, in separate. Under these parameters, there are five possible fixed points on the boundary and an equilibrium curve for the system in total. Among the five boundary fixed points, $(x=0, r_c=1.5)$ is always stable and $(x=0, r_c=3.5)$, $(x=1, r_c=1.5)$, $(x=1, r_c=3.5)$ are always unstable, while the stability of $(x=1, r_c=\frac{25}{16}+\frac{a_i}{2})$ depends on the parameter $a_i$. Detailed proofs are provided in \emph{Appendix A}. Of particular interest, here we focus on the stability of the equilibrium curve. Unusually, we obtain the emergence of stable equilibrium manifolds in both Fig. \ref{2}(B) and Fig. \ref{2}(C), which means the system can finally evolve to many different stable states depending on the initial conditions. This new phenomenon is totally different from the solely interior fixed points situation discussed in previous models~\cite{weitz2016oscillating, tilman2019evolutionary}. Moreover, the equilibrium curve of unstable manifold situations in our model (see Fig. \ref{2}(A)) is in line with the phase separatrix observed by experimental work in Fig. \ref{exp}(A), which indicates the potential power for our general framework to explain the abundant eco-evolutionary phenomena shown in real microbial systems. Additionally, we provide detailed proof of the stability of the manifolds in the following section. For simplicity and without loss of generality, we use parameters given in Fig. \ref{2}(C) in which $\theta_1=2$ and $a_1=0$ as an example. The schematic of this proof is shown in Fig. \ref{proof}.

Assume $(x^*, r_c^*)$ is an arbitrary fixed point on the equilibrium curve $P_c=P_d$. The phase plane is separated into two regions by the equilibrium curve, over which is named as region $A$ while beneath which is region $B$. Then $(x^*, r_c^*)$ is stable if and only if 
\begin{equation}
\vec{w_i} \cdot \vec{v_i} < 0
\label{eproof1}
\end{equation}
in a small neighborhood of $(x^*, r_c^*)$, where $\vec{v_i}$ is the normal vector of the equilibrium curve while $\vec{w_i}$ is the trajectory field direction, $i=1, 2$.  According to Eq. \ref{e2}, the equilibrium curve $P_c=P_d$ can be written as
\begin{equation}
r_c=r_d+\frac{S+1-r_d}{Sx+1}.
\label{e4}
\end{equation}
Define $h(x)$ as the slope equation of the equilibrium curve, we have
\begin{equation}
h(x)=\frac{\mathrm{d}r_c}{\mathrm{d}x}=-\frac{S(S+1-r_d)}{(Sx+1)^2}
\label{e3}
\end{equation}
Therefore we get the normal vectors of the equilibrium curve at $(x^*, r_c^*)$:
\begin{equation}
\begin{split}
\vec{v_1}&=(-h(x^*), 1)\\
\vec{v_2}&=(h(x^*), -1).
 \end{split}
 \label{eproof2}
\end{equation}
On the other hand, the trajectory vector $(\dot{x}, \dot{r_c})$ satisfies $\dot{r_c}<0, \dot{x}>0$ when $(x, r_c)$ is in region $A$ and $\dot{r_c}>0, \dot{x}<0$ when $(x, r_c)$ is in region $B$, according to Eq. \ref{e1}. Thus the trajectory filed directions in a small neighborhood of $(x^*, r_c^*)$ read
\begin{equation}
\begin{split}
\vec{w_1}&=(1, \lim_{P_c\to P_d} g(x^*)), P_c \to P_d+\delta\\
\vec{w_2}&=(-1, -\lim_{P_c\to P_d} g(x^*)), P_c \to P_d-\delta,
 \end{split}
 \label{eproof3}
\end{equation}
where $g(x)$ is the slope function of the trajectory field:
\begin{equation}
\begin{split}
	g(x)&=\frac{\mathrm{d}r_c}{\mathrm{d}x} = \frac{\mathrm{d}r_c/\mathrm{d}t}{\mathrm{d}x/\mathrm{d}t}\\
	      &=\frac{\epsilon (r_c-\alpha)(\beta-r_c)(P_c-P_d)(a_1-(\theta_1 P_c-P_d))}{x(1-x)(P_c-P_d)}
 \end{split}
 \label{eproof4}
\end{equation}
Substituting Eq. \ref{eproof2} and Eq. \ref{eproof3} in Eq. \ref{eproof1}, we reach to the necessary and sufficient condition that $(x^*, r_c^*)$ is stable: 
\begin{equation}
\lim_{P_c\to P_d}g(x^*)<h(x^*)
\label{eproof5}
\end{equation}


Finally let $\alpha=1.5$, $\beta=3.5$, $S=3$,  $r_d=1.5$, $\epsilon=2$, $\theta_1=2$ and $a_1=0$, we have
\begin{equation}
\begin{split}
        h(x^*)&=-\frac{7.5}{(3x^*+1)^2}\\
	\lim_{P_c\to P_d} g(x^*)&=\frac{540x^* - 45}{16(3x^* + 1)^2(x^* - 1)}	                                
 \end{split}
 \label{eproof6}
\end{equation}
Note that $r_c\in [1.5, 3.5]$, which leads to $x\in [1/12, 1]$ according to Eq. \ref{e4}. Substituting Eq. \ref{eproof6} to Eq. \ref{eproof5}, we have $x^*>0.25$. Therefore when $x>0.25$, the trajectories in the neighborhood of the fixed points will converge to the equilibrium curve, which proves the stability of the manifolds. On the contrary, when $x\in [1/12, 0.25)$, we have $\vec{w_i} \cdot \vec{v_i} >0$, the trajectories in the neighborhood evolves away from the equilibrium curve, in which situation the manifolds are unstable. Therefore, $x=0.25$ is actually a saddle point of the system which satisfies $\vec{w_i} \cdot \vec{v_i} =0$. The typical stable and unstable manifold situations in a small neighborhood of the equilibrium curve are marked with red outlines in Fig. \ref{2}(C). Similarly, we derive that the stable region of the equilibrium curve is $x\in (0.5595,1)$ in Fig. \ref{2}(B), while there only exists unstable manifolds in Fig. \ref{2}(A). In summary, we have proved the emergence of stable equilibrium manifolds in our framework. Meanwhile, we provided a detailed approach to calculate the saddle point of the system, which is the critical point for the stability of the equilibrium curve. 

\begin{figure*}[htbp]
    \centering
    \includegraphics[width=0.95\linewidth]{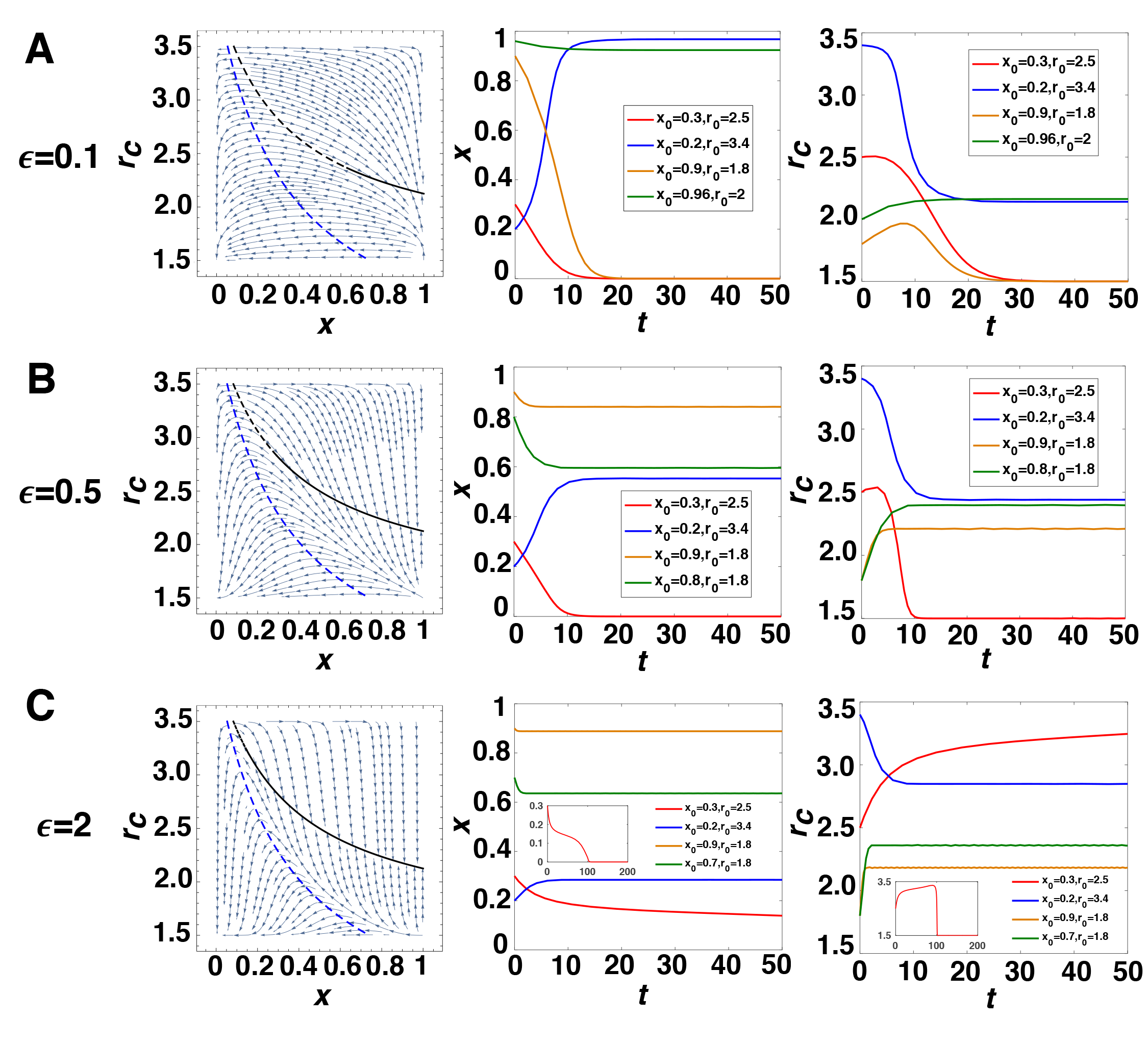}
    \caption{The effects of relative asymmetrical feedback speed. In all subfigures,  $\alpha=1.5$, $\beta=3.5$, $S=3$,  $r_d=1.5$, $\theta_1=4$, $a_1=0$. We change $\epsilon=0.1, 0.5, 2$ in (A)-(C) respectively. The first column presents phase graphs in different situations while the second and third columns give variations of strategy dynamics as well as the cooperator's multiplication factor over time, correspondingly. We use solid vs dashed lines to indicate the stability of equilibrium curve in all phase graphs.}
    \label{3}
\end{figure*}

In Fig. \ref{3}, we present the detailed effects of relative asymmetrical feedback speed. In all subfigures, $\alpha=1.5$, $\beta=3.5$, $S=3$,  $r_d=1.5$, $\theta_1=4$, $a_1=0$. We give $\epsilon=0.1, 0.5, 2$ in Fig. \ref{3}(A)-(C) separately. According to the results presented by the first column which shows phase graphs under different circumstances, we find that the relative changing speed of cooperator's multiplication factor has a strong influence on the slope of trajectories and further affects the position of saddle point on equilibrium curve, i.e., influences the stability of the manifolds. When the relative feedback speed $\epsilon$ is larger, the trajectory field acts steeper and the attraction basin of these stable manifolds increases. We prove this conclusion analytically using the same approach as shown in the stability proof section. We keep $\epsilon$ as a variable and the other parameters are the same as in Fig. \ref{3}:

Restraining the slope function of trajectory field $g(x)$ in a small neighborhood of the equilibrium curve, i.e., letting $P_c\to P_d$, leads to $r_c \to 1.5+2.5/(3x+1)$ and $x\in [1/12, 1]$: 
\begin{equation}
\begin{split}
\lim_{P_c\to P_d} g(x)&=\frac{\epsilon (r_c-\alpha)(\beta-r_c)(a_1-(\theta_1 P_c-P_d))}{x(1-x)}\\
	                                   &=\frac{135 \epsilon (12x-1)}{32(3x + 1)^2 (x - 1)}
\end{split}
\end{equation}
Let $\lim_{P_c\to P_d}g(x)=h(x)$, we have 
\begin{equation}
\begin{split}
	x^*(\epsilon)&=\frac{27\epsilon + 48}{324\epsilon+48}\\
	 &=\frac{1}{12}+\frac{44}{324\epsilon+48}
\end{split}
\end{equation}
Therefore when $x> x^*$ the equilibrium curve is stable, while when $x\in [1/12, x^*)$ the equilibrium curve is unstable. Finally, note that $x^*(\epsilon)$ is a monotone decreasing function which indicates when $\epsilon$ is larger, $x^*$  becomes smaller, i.e, the attraction basin of these stable manifolds increases.  

In addition, the second and third column of Fig. \ref{3} give variations of strategy dynamics and the cooperator's multiplication factor over time, respectively. In each subfigure, we present four different initial conditions. Results show that a larger $\epsilon$ effectively accelerates the convergence speed of stable manifold, while on the hand may slow down the converging process of unstable manifolds towards mutual defection state. 

\begin{figure}[htbp]
    \centering
    \includegraphics[width=0.95\linewidth]{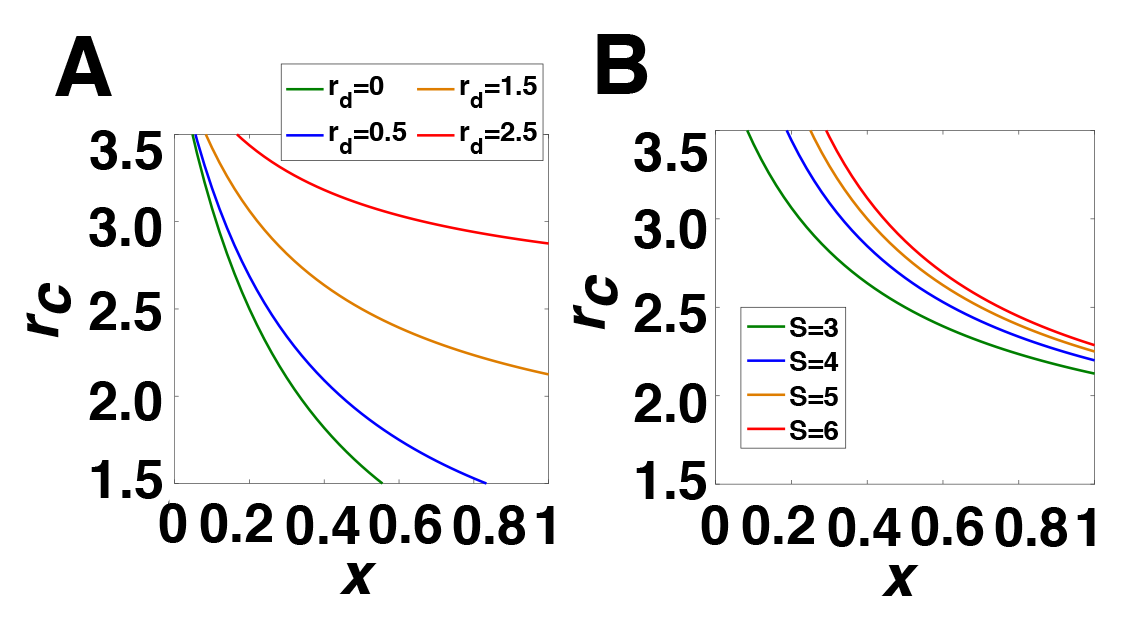}
    \caption{How defector's multiplication factor $r_d$ and the group size $S$ influence the position of equilibrium curve and stable manifolds. (A) We fix $S=3$ and change $r_d=0, 0.5, 1.5, 2.5$ respectively. (B) We fix $r_d=1.5$ and set $S=3, 4, 5, 6$ respectively. }
    \label{4}
\end{figure}

In Fig. \ref{4},  we show the determinants of the position of equilibrium curve as well as their detailed effects. In general framework, the equilibrium equation $r_c = r_d+\frac{S+1-r_d}{Sx+1}$ indicates that the position of equilibrium curve is determined by defector's multiplication factor $r_d$ and the group size $S$, according to Eq. \ref{e4}. Moreover, we have
\begin{equation}
	\begin{split}
		\frac{\partial{r_c}}{\partial{r_d}} &=\frac{Sx}{Sx+1} \\
		\frac{\partial{r_c}}{\partial{S}} &= \frac{1-(1-r_d)x}{(Sx+1)^2}\\
	\end{split}
\end{equation}
in which $r_d\ge 0$ and $x\in [0, 1]$. Therefore $\frac{\partial{r_c}}{\partial{r_d}}\ge 0$, $\frac{\partial{r_c}}{\partial{S}}\ge 0$, which means $r_c$ increases as $r_d$ or $S$ becomes larger. We present these variation trends in  Fig. \ref{4} in detail, in which $S=3$ and $r_d=0, 0.5, 1.5, 2.5$ in Fig. \ref{4}(A) while $r_d=1.5$ and $S=3, 4, 5, 6$ in Fig. \ref{4}(B), respectively.

\begin{figure*}[htbp]
    \centering
    \includegraphics[width=0.95\linewidth]{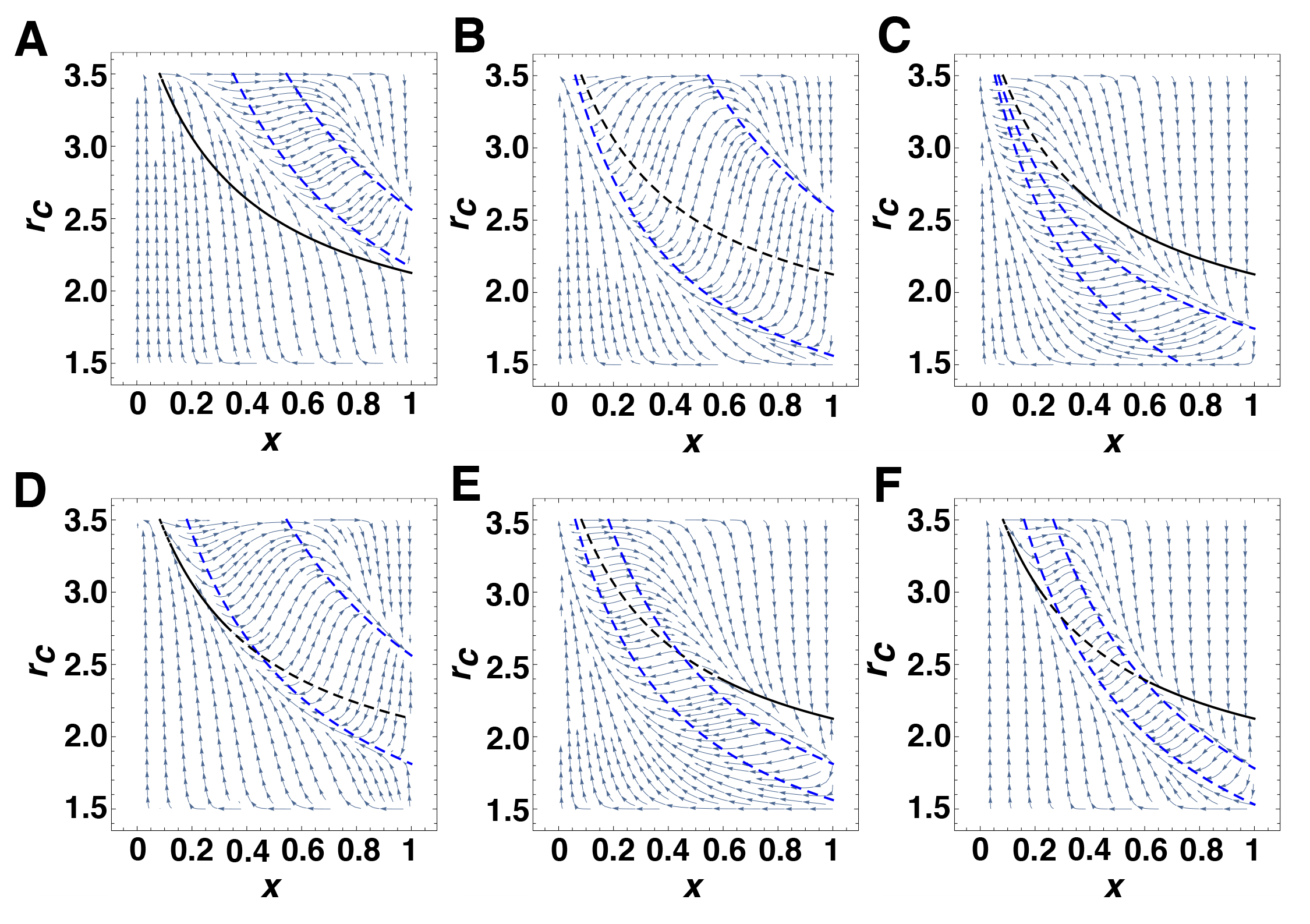}
    \caption{Phase graphs under different circumstances where the control function $f$ is in cubic forms. In all subfigures, $\alpha=1.5$, $\beta=3.5$, $S=3$,  $r_d=1.5$, $\epsilon=2$. We use solid vs dashed lines to indicate the stability of equilibrium curves. (A) $\theta_1=2$, $a_1=1.2$, $\theta_2=2$, $a_2=2$. The stable range of the equilibrium curve is $x\in (0.09355,1]$. (B) $\theta_1=2$, $a_1=0$, $\theta_2=2$, $a_2=2$. There only exists unstable manifolds. (C) $\theta_1=1.5$, $a_1=0$, $\theta_2=4$, $a_2=0$. The stable range of the equilibrium curve is $x\in (0.3395,1]$. (D) $\theta_1=2$, $a_1=0.5$, $\theta_2=2$, $a_2=2$. The stable range of the equilibrium curve is $x\in (0.1237,0.3198]$. (E) $\theta_1=2$, $a_1=0$, $\theta_2=2$, $a_2=0.5$. The stable range of the equilibrium curve is $x\in (0.5984,1]$. (F) $\theta_1=4$, $a_1=1$, $\theta_2=4$, $a_2=2$. The stable range of the equilibrium curve is $x\in (0.1057,0.2226)$ and $x\in (0.622,1]$. }
    \label{5}
\end{figure*}

Furthermore, in Fig. \ref{5}, we propose a more complex situation where $n=2$ and $f$ is a cubic function, which writes
 \begin{equation}
	f(x, r_c) = (P_c-P_d)((\theta_1 P_c-P_d)-a_1)(a_2-(\theta_2 P_c-P_d))
\end{equation}
In all subfigures, $\alpha=1.5$, $\beta=3.5$, $S=3$,  $r_d=1.5$, $\epsilon=2$. In Fig. \ref{5}(A)-(C), we present different combinations of the two control curves when there is no intersection between control curves and equilibrium curve, while there exists one intersection in Fig. \ref{5}(D)-(E),  and finally two intersections in Fig. \ref{5}(F). Not surprisingly, the trajectory field reveals rich probabilities and abundant new variation trends, an important one of which is $(x=0, r_c=3.5)$ becomes a stable fixed point while $(x=0, r_c=1.5)$ becomes unstable, indicating an opposite path direction for a range of initial conditions compared to the $n=1$ situations we discussed above. In addition, in Fig. \ref{5}(A) and Fig. \ref{5}(C)-(F), we obtain existence of stable equilibrium manifolds ranging from a small neighborhood of the equilibrium curve like in Fig. \ref{5}(D), to a large region like in Fig. \ref{5}(A), as shown by solid lines. On the other hand, Fig. \ref{5}(B) shows completely unstable manifolds situation. 

In general, our results reveal that we can control the stability of the equilibrium curve $P_c=P_d$ as well as the stability of the manifolds by giving different feedback control functions $f$ or changing the relative feedback speed $\epsilon$. This indicates the possibility for designing population-state dependent switching control laws to steer the system evolution towards desired directions using our framework, which may have many applications in microbial experiments, such as designing culture refresh modes and building needed chemostat. We provide detailed control approach as well as two control examples in the next section.

\subsection{Manifold control with switching control laws}
Firstly, we incorporate two kinds of switching control laws in our general framework, one of which is time-dependent and the other one is state-dependent.

Time-dependent switching control law for the control function $f$ can be written as follows:
\begin{equation}
\begin{split}
	f_{\sigma(t)}(x, r_c) &= f_{\sigma, k(\sigma)}(x, r_c) \\
	                         &= (P_c-P_d)((\theta_{\sigma, 1} P_c-P_d)-a_{\sigma, 1})\\
	                         &...(a_{\sigma, k(\sigma)}-(\theta_{\sigma, k(\sigma)} P_c-P_d))
\end{split}
\end{equation}
Here $\sigma(t): [0,\infty) \rightarrow I$, where $I$ represents a finite index set ${1,2,...,m}$. In addition, $k(\sigma)+1$ is the order of control function $f_{\sigma}$. 

Additionally, state-dependent switching control law can be written as: 
\begin{equation}
\begin{split}
	f_{s}(x, r_c) &= f_{s, k(s)}(x, r_c) \\
	            &= (P_c-P_d)((\theta_{s, 1} P_c-P_d)-a_{s, 1})\\
	            &...(a_{s, k(s)}-(\theta_{s, k(s)} P_c-P_d))
\end{split}
\end{equation}
Here $s: q\rightarrow I$, where $q \in Q$, which represents a family of switching surfaces/guards that partition the $x-r_c$ plant into finite regions, and $I$ is a finite index set ${1,2,...,m}$. $k(s)+1$ is the order of control function $f_{s}$.

\begin{figure*}[htbp]
    \centering
    \includegraphics[width=0.95\linewidth]{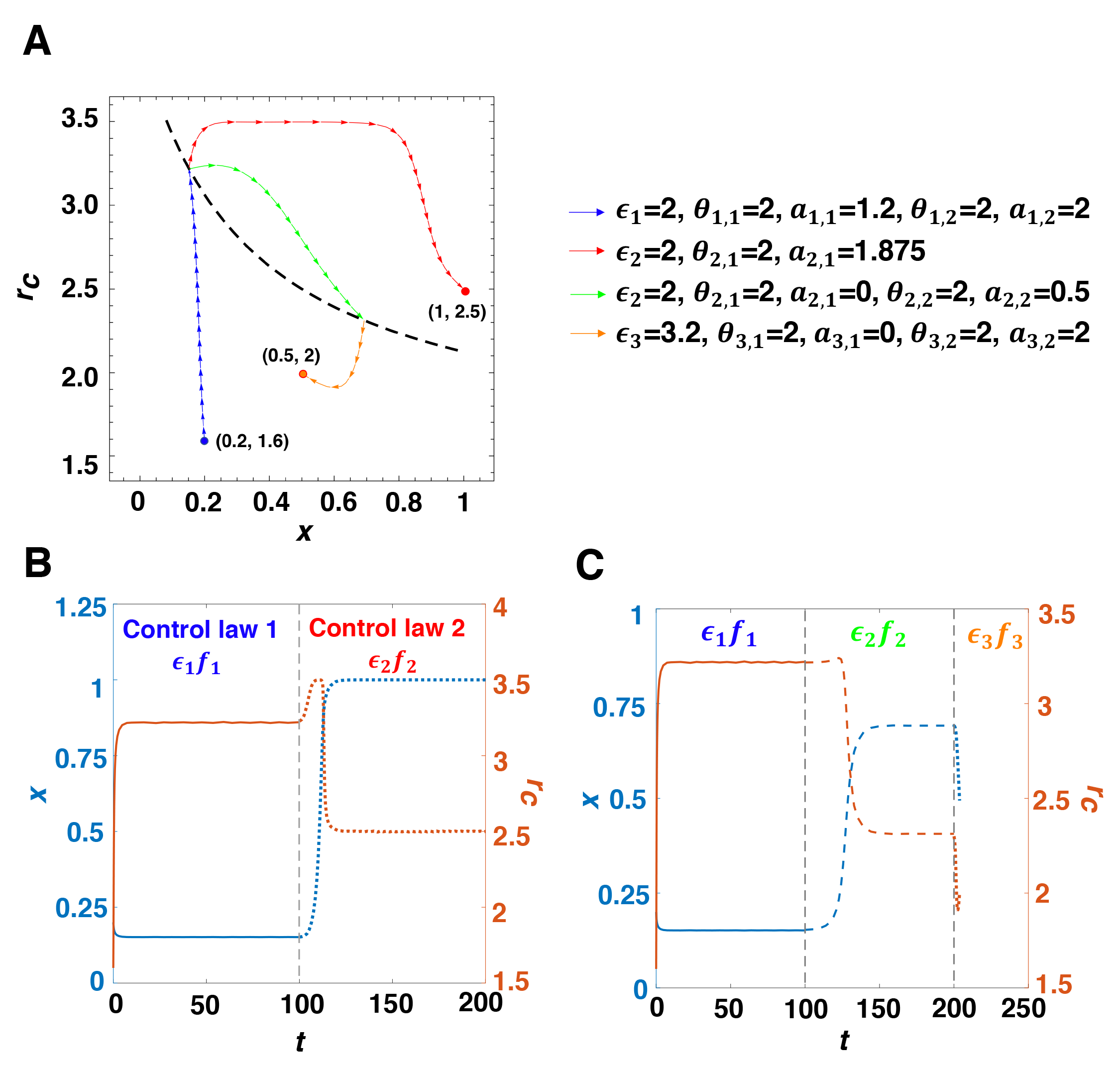}
    \caption{Two control examples with switching control laws in general framework. The fixed parameters are as follows: $\alpha=1.5, \beta=3.5, S=3, r_d=1.5$. (A) Phase trajectories of two control examples. We begin from $(x_0, r_0)=(0.2,1.6)$ and give two desired final state: $(x_1, r_1)=(1, 2.5)$ and $(0.5, 2)$. The first evolution path is the blue trajectory followed by the red one, ending at (1, 2.5). Here we only use time-dependent switching control law, in which $\epsilon_1=2$ with $\theta_{1,1}=2, a_{1,1}=1.2, \theta_{1,2}=2, a_{1,2}=2$ for $f_1$ and $\epsilon_2=2$ with $\theta_{2,1}=2, a_{2,1}=1.875$ for $f_2$. Here $t$ is large enough for ending the evolutions at the stable fixed points. The second example is consist of three trajectories: the blue, green and orange one, ending at $(0.5, 2)$. Here we use time-dependent switching control law for the first two trajectories, and state-dependent switching control law for the third control function to stop the evolution at $x=0.5$. Detailed parameters for the relative feedback speeds and the control functions are as follows: $\epsilon_1=2$ with $\theta_{1,1}=2, a_{1,1}=1.2, \theta_{1,2}=2, a_{1,2}=2$ for $f_1$, $\epsilon_2=2$ with $\theta_{2,1}=2, a_{2,1}=0, \theta_{2,2}=2, a_{2,2}=0.5$ for $f_2$, and $\epsilon_3=3.2$ with $\theta_{3,1}=2, a_{3,1}=0, \theta_{3,2}=2, a_{3,2}=2$ for $f_3$. In both examples, when system evolves to the equilibrium curve and reaches to a stable fixed point, we change the control function and meanwhile provide a small disturbance $\delta \le 0.01$ to $r_c$ according to the desired direction, in which way to restart the evolutions. (B)(C) Detailed time evolutions of the population state $(x, r_c)$ in the first and second control example, respectively. The colors of the curves are corresponding to two y-Axes while the types of the curves indicate evolutions under different control laws. The colors of control laws $\epsilon_i f_i$ are corresponding to the trajectories in (A). Specially, In (C), the control law 3 is a state-dependent switching control law and the evolution ends at about $t=203.8$. }
    \label{6}
\end{figure*}

In \emph{Appendix B}, we provide a constructive proof for the existence of control laws when given a certain final state $(x_1, r_1)$ with an initial state $(x_0, r_0)$ in our general framework. We show the possibility of ending the evolution in any desired region, both over or beneath the equilibrium curve, with any fraction of final cooperators, as long as the initial fraction of cooperators $x_0$ is not too small, otherwise the evolution even cannot reach to the equilibrium curve and directly ends into a mutual defection state. Note that our proof is a heuristic method for seeking a group of possible control functions, which might not be the only solution for a certain control problem. 

For a better understanding, here we illustrate two detailed control examples in Fig. \ref{6}. The evolution begins from $(x_0, r_0)=(0.2,1.6)$ and we give two desired final states: $(x_1, r_1)=(1, 2.5)$ and $(0.5, 2)$. 

The first example only use time-dependent switching control law, in which the control functions and the relative feedback speeds are as follows:
\begin{equation}
\begin{split}
	\epsilon_1 f_{1} &= \epsilon_1(P_c-P_d)((\theta_{1, 1} P_c-P_d)-a_{1, 1})\\
	                      &*(a_{1, 2}-(\theta_{1, 2} P_c-P_d))\\
	        &=2(P_c-P_d)((2P_c-P_d)-1.2)(2-(2P_c-P_d))\\
	\epsilon_2 f_{2} &= \epsilon_2(P_c-P_d)(a_{2, 1}-(\theta_{2, 1} P_c-P_d))\\
	        &=2(P_c-P_d)(1.875-(2P_c-P_d))
\end{split}
\end{equation}
with $t$ large enough for ending the evolutions at the stable fixed points. The evolution path is the blue trajectory followed by the red one, which finally stops at $(1, 2.5)$, as shown in Fig. \ref{6}(A). Further, in Fig. \ref{6}(B), we provide detailed time evolutions of the population state $(x, r_c)$ under two control stages. The colors of the curves are corresponding to two y-Axes while the types of the curves indicate evolutions under different control laws. The colors of control laws $\epsilon_i f_i$ are corresponding to the trajectories in Fig. \ref{6}(A). 

The second example is composed of three trajectories in Fig. \ref{6}(A): the blue, green and orange one, ending at $(0.5, 2)$. Here we use time-dependent switching control law for the first two trajectories, and state-dependent switching control law for the third control function to stop the evolution at $x=0.5$. The control functions and the relative feedback speeds are as follows:
\begin{equation}
\begin{split}
	\epsilon_1 f_{1} &= 2(P_c-P_d)((2P_c-P_d)-1.2)(2-(2P_c-P_d))\\
	\epsilon_2 f_{2} &= 2(P_c-P_d)((2P_c-P_d)-0)(0.5-(2P_c-P_d))\\
	\epsilon_3 f_{3} &= 3.2(P_c-P_d)((2P_c-P_d)-0)(2-(2P_c-P_d))\\
\end{split}
\end{equation}
Additionally, in Fig. \ref{6}(C), we show time evolutions of the population state $(x, r_c)$ under three control stages. Specially, the control law $3$ is state-dependent and the evolution stops at about $t=203.8$.

In both examples, when system evolves to the equilibrium curve and reaches to a stable fixed point, we change the control function and meanwhile provide a small disturbance $\delta \le 0.01$ to $r_c$ according to the desired direction, in which way to restart the evolutions. The absolute errors for the final positions of $r_c$ when $x$ reaches to $x_1$ is within $0.01$, according to the numerical solutions of the equations.

\section{Conclusions And Discussions}
The coevolutionary game between environment and strategy dynamics has aroused great concern in recent years, in order to account for the widespread existence of eco-evolutionary feedback loops and their great importance in a range of natural systems. The existing feedback-evolving models almost exclusively concentrate on two-player games with linear environmental feedback laws. However, there still lacks of a general framework to describe the coevolutionary dynamics of population strategy and population properties in PGG. Meanwhile, the detailed effects of nonlinearity in feedback loops remain unclear. Moreover, how to effectively steer the coevolutionary games dynamics to desired population states with external feedback control laws in microbial experiments is a challenging theoretical gap.

Inspired by the experimental results in Ref.~\cite{sanchez2013feedback} that confirm the existence of strong feedback loops as well as the emergence of separatrix line in dynamics, we provide a general framework in multi-player games with asymmetrical feedback driven by nonlinear selection gradient, taking into account the fact that the driving factor for the existence of feedback loops in most microbial systems is the preferential access to the common good for cooperators. We consider ecological properties solely being affected by the fraction of cooperators, such as population density in ~\cite{sanchez2013feedback}, as a benefit of cooperation that can be described by cooperator's multiplication factor, which makes our framework more general. Unusually, we find the emergence of multiple segments of stable and unstable equilibrium manifolds in phase graphs with a number of control functions, which is a new dynamical phenomenon that totally different from the solely interior fixed equilibrium situation obtained in previous works. In addition, we show that the relative asymmetrical feedback speed for group interactions centered on cooperators can accelerate the converging process of the stable manifolds while slow down the convergence speed of unstable manifolds. And a larger feedback speed increases the attraction basin of these stable manifolds. Besides, the position of the equilibrium curve is determined by defector's multiplication factor and the group size of the game. Finally, we propose an innovative manifold control approach by incorporating switching control laws which can actually control the stability of the possible manifolds into our general framework. We provide a constructive proof as well as two specific control examples to show the possibility of steering the system states evolving towards designed directions and entering into any desired region, as long as the initial fraction of cooperators is not too small.   

Our work sheds light on the potential effects of the asymmetrical feedback driven by nonlinear selection gradient in PGG, which is an important generalization and complement for the current coevolutionary game theory frameworks. While some previous works study the stochastic dynamics on slow manifolds in deterministic dynamical systems~\cite{constable2013stochastic, constable2016demographic}, here for the first time, we find the emergence of stable equilibrium manifolds in feedback-evolving games. In light of the extensive existence of feedback loops related to common goods in microbial systems as well as the consistency of unstable equilibrium manifold in our general framework and the phase separatrix in experimental results, our model reveals great potentials to explain and understand the eco-evolutionary dynamics in a variety of real microbial systems. Furthermore, the manifold control method that can steer eco-evolutionary dynamics with external switching feedback control laws may have wide applications in microbial experiments, such as designing culture refresh modes for continuous culture devices and establishing needed chemostat, which is of great significance in both systems biology and microbial ecology.

\section*{Acknowledgement}
X.W. \& Z.Z. gratefully acknowledge Alvaro Sanchez who generously shared the experimental data for our work. This work is supported by Program of National Natural Science Foundation of China Grant No. 11871004, Fundamental Research of Civil Aircraft Grant No. MJ-F-2012-04. F.F. is supported by a Junior Faculty Fellowship awarded by the Dean of the Faculty at Dartmouth and also by the Dartmouth Faculty Startup Fund, the Neukom CompX Faculty Grant, and Walter \& Constance Burke Research Initiation Award.


\clearpage 

\end{document}


\newcommand{\beginsupplement}{%
        \setcounter{table}{0}
        \renewcommand{\thetable}{S\arabic{table}}%
        \setcounter{figure}{0}
        \renewcommand{\thefigure}{S\arabic{figure}}%
     }
     
\beginsupplement     

\section*{Appendix A: Stability of boundary fixed points when the feedback control function $f$ is in quadratic forms. }

Here we discuss the stability of all possible fixed points on the boundary  (not include the intersection of equilibrium curve and the boundary, which is considered as a part of equilibrium curve) when the feedback control $f$ is in quadratic forms, using parameters shown in \textbf{Fig. 2} in the main text. The co-evolutionary model can be written as 
\begin{equation}
	\begin{cases}
	\begin{split}
		\dot{x} &= x(1-x)\left(P_c-P_d\right)\\
		\dot{r_c} &= \epsilon (r_c-\alpha)(\beta-r_c)(P_c-P_d)(a_1-(\theta_1 P_c-P_d))
	\end{split}
	\end{cases}
	\label{e1}
\end{equation}
in which 
\begin{equation}
    \begin{split}
        P_c &= \frac{1+Sx}{S+1}r_c -1\\
        P_d &= \frac{Sx}{S+1}r_d
        \label{e2}
    \end{split}
\end{equation}
Applying Eq. \ref{e2} as well as the fixed parameters $\alpha=1.5$, $\beta=3.5$, $S=3$,  $r_d=1.5$, $\epsilon=2$, $\theta_1=2$ into Eq. \ref{e1}, we have
\begin{equation}
	\begin{cases}
	\begin{split}
		\dot{x} &= x(1-x)(\frac{1+3x}{4}r_c-1-\frac{4.5}{4}x)\\
		\dot{r_c} &=2(r_c-1.5)(3.5-r_c)(\frac{1+3x}{4}r_c-1-\frac{4.5}{4}x)(a_1-(2(\frac{1+3x}{4}r_c-1)-\frac{4.5}{4}x))
	\end{split}
	\end{cases}
	\label{e1}
\end{equation}
Denote $J$ as the Jacobian matrix of the system. We have five boundary fixed points in total: 
\normalsize{}

(1) $x=0, r_c=1.5$
\begin{equation*}
    J(x=0, r_c=1.5) = \left[\begin{matrix}
    -\frac{5}{8} & 0\\
    0 & -\frac{5a_1}{2}- \frac{25}{8}
    \end{matrix}\right]
\end{equation*}

The eigenvalues of the Jacobian matrix are \(\lambda_1=-\frac{5}{8}<0\) and \(\lambda_2= -\frac{5a_1}{2}- \frac{25}{8}<0\), since $a_1>0$ in our framework. Therefore the fixed point $(x=0, r_c=1.5)$ is always stable.

(2)$x=0, r_c=3.5$.
\begin{equation*}
    J(x=0, r_c=3.5) = \left[\begin{matrix}
    -\frac{1}{8} & 0\\
    0 & \frac{a_1}{2}+ \frac{1}{8}
    \end{matrix}\right]
\end{equation*}

The eigenvalues of the Jacobian matrix are \(\lambda_1=-\frac{1}{8}<0\) and \(\lambda_2= \frac{a_1}{2}+\frac{1}{8}>0\). Therefore the fixed point $(x=0, r_c=3.5)$ is always unstable.

(3) $x=1, r_c=1.5$.
\begin{equation*}
    J(x=1, r_c=1.5) = \left[\begin{matrix}
    \frac{5}{8} & 0\\
    0 & -\frac{5a_1}{2}- \frac{5}{16}
    \end{matrix}\right]
\end{equation*}

The eigenvalues of the Jacobian matrix are \(\lambda_1=\frac{5}{8}>0\) and \(\lambda_2= -\frac{5a_1}{2}- \frac{5}{16}<0\). Therefore the fixed point $(x=1, r_c=1.5)$ is always unstable.

(4) $x=1, r_c=3.5$.
\begin{equation*}
    J(x=1, r_c=3.5) = \left[\begin{matrix}
    -\frac{11}{8} & 0\\
    0 & \frac{341}{16}- \frac{11a_1}{2}
    \end{matrix}\right]
\end{equation*}

The eigenvalues of the Jacobian matrix are \(\lambda_1=-\frac{11}{8}<0\) and \(\lambda_2= \frac{341}{16}- \frac{11a_1}{2}\). In the main text, we change $a_1=2, 0.5, 0$ respectively. Under all these circumstances, $\lambda_2>0$. Therefore the fixed point $(x=1, r_c=3.5)$ is always unstable.

(5) $x=1, r_c=\frac{25}{16}+\frac{a_1}{2}$, which is the intersection of the control curve $\theta_i P_c-P_d=a_i$ and the boundary $x=1$.
\begin{equation*}
    J(x=1, r_c=\frac{25}{16}+\frac{a_1}{2}) = \left[\begin{matrix}
    \frac{9}{16}-\frac{a_1}{2} & 0\\
 (\frac{a_1}{2}-\frac{9}{16})(\frac{a_1}{2}-\frac{31}{16})(a_1+\frac{1}{8})(\frac{3a_1}{4}+\frac{39}{32}) & 2(\frac{a_1}{2}-\frac{9}{16})(\frac{a_1}{2}-\frac{31}{16})(a_1+\frac{1}{8})    \end{matrix}\right]
\end{equation*}

The eigenvalues of the Jacobian matrix are \(\lambda_1= \frac{9}{16}-\frac{a_1}{2}\) and \(\lambda_2= 2(\frac{a_1}{2}-\frac{9}{16})(\frac{a_1}{2}-\frac{31}{16})(a_1+\frac{1}{8})\). Therefore, when $a_1\in (\frac{9}{8}, \frac{31}{8})$, we have $\lambda_1<0$ and $\lambda_2<0$ at the same time, the fixed point $(x=1, r_c=\frac{25}{16}+\frac{a_1}{2})$ is stable. When $a_1\in [0, \frac{9}{8})$ or $a_1>\frac{31}{8}$, however, the fixed point is unstable.

\normalsize{}
In conclusion, we analyze the stability of all five boundary fixed points in \textbf{Fig. 2} in the main text, among which $(x=0, r_c=1.5)$ is always stable and $(x=0, r_c=3.5)$, $(x=1, r_c=1.5)$, $(x=1, r_c=3.5)$ are always unstable, while the stability of $(x=1, r_c=\frac{25}{16}+\frac{a_i}{2})$ depends on the parameter $a_i$.

\section*{Appendix B: Constructive proof for the existence of control laws in the general framework. }

\begin{figure*}[!h]
    \centering
    \includegraphics[width=0.95\linewidth]{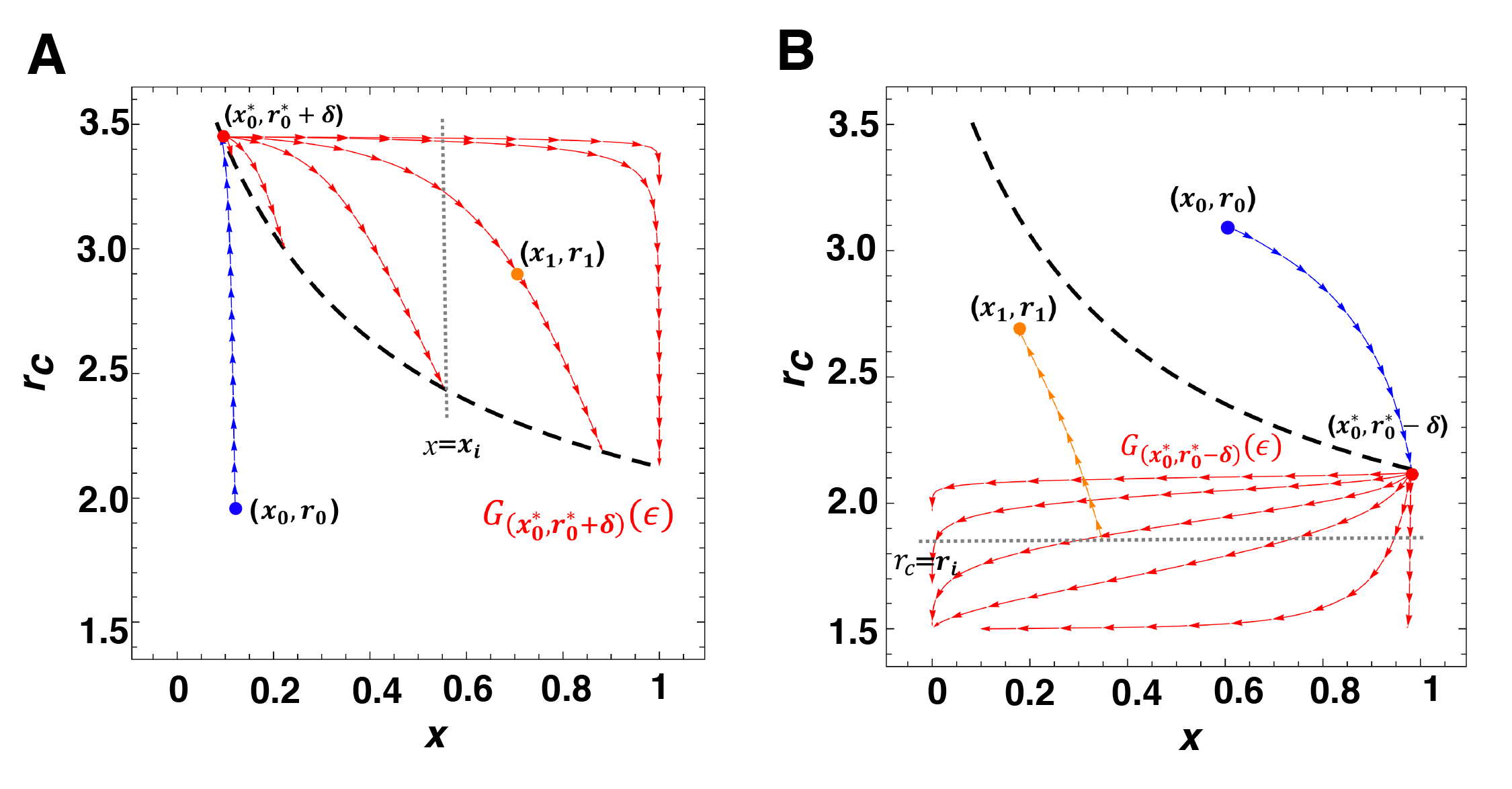}
    \caption{The existence of control laws for any desired final state $(x_0, r_0)$ with a given initial state $(x_1, r_1)$. (A)  $(x_0, r_0)$ is beneath the equilibrium curve while $(x_1, r_1)$ is over the equilibrium curve. (B) $(x_0, r_0)$ is over the equilibrium curve while $(x_1, r_1)$ is beneath the equilibrium curve.}
    \label{S1}
\end{figure*}

Here we provide a constructive proof for the existence of control laws when given a certain final state $(x_1, r_1)$ with an initial state $(x_0, r_0)$. For simplicity and without loss of generality, we fix the following parameters: $\alpha=1.5, \beta=3.5, S=3, r_d=1.5$ in our general framework. 

(i) In \textbf{Fig. S1A}, we prove the existence of control laws when $(x_0, r_0)$ is beneath the equilibrium curve while $(x_1, r_1)$ is over the equilibrium curve. Under the fixed parameters, when $x_0 \le 1/12$, the initial fraction of cooperators is too small that the system will finally evolve into a mutual defection state in our framework. Assume $x_0>1/12$. (a) Firstly, we prove that we can control any given initial state evolving to the stable fixed point on the equilibrium curve that can arbitrarily approaches to the endpoint $(1/12, 3.5)$. When $x_0$ is close to $1/12$, we can simply use $f_1=(P_c-P_d)((2P_c-P_d)-1.2)(2-(2P_c-P_d))$ with time-dependent switching control law. Let $\epsilon$ be sufficiently large, then $(x_0, r_0)$ will finally stop at a stable fixed point $(x^*_0, r^*_0)$ which can arbitrarily approaches to $(1/12, 3.5)$ on the equilibrium curve, like shown in the figure. When $x_0$ is relatively large, we may use state-dependent control laws in which $f_0=(P_c-P_d)(2-(2P_c-P_d))$ and change $\epsilon$ to control the initial state evolving to a desired new region where $x_0$ is close to $1/12$ in advance. Then similarly, we use $f_1$ and let $\epsilon$ be large enough to control the system evolving to the stable fixed point $(x^*_0, r^*_0)$. (b) Then we prove that the trajectories begin from $(x^*_0, r^*_0+\delta)$ which is close to $(1/12, 3.5)$ can actually cover the whole region over the equilibrium curve. Here $\delta$ denotes a small disturbance on $r^*_0$. We present a possible control function $f_2=(P_c-P_d)(0-(4P_c-P_d))$. Define $G_{(x^*_0,r^*_0+\delta)}(\epsilon)$ as the set of trajectories that begin from $(x^*_0,r^*_0+\delta)$ with control function $f_2$ and relative feedback speed $\epsilon$. $G_{(x^*_0,r^*_0+\delta)}(\epsilon)$ is a continuous function of $t, x, r_c$ and $\epsilon$. When $\epsilon$ is small enough, $G_{(x^*_0,r^*_0+\delta)}(\epsilon)$ finally stops at $(1, 2.125)$ which is the endpoint of the equilibrium curve. Meanwhile, when $\epsilon \to \infty$, the trajectory stops at a stable fixed point that infinitely approaches to $(x^*_0, r^*_0)$. According to the continuity of $\epsilon$, we derive that the trajectory set $G_{(x^*_0,r^*_0+\delta)}(\epsilon)$ can end at any fixed point ranging from $(x^*_0, r^*_0)$ to $(1, 2.125)$. Further, given any $x=x_i$, when $\epsilon \to 0$, $G_{(x^*_0,r^*_0+\delta)}(0)$ goes through $(x_i, r^*_0)$ while we already proved that there exists a $\epsilon_i$ that makes $G_{(x^*_0,r^*_0+\delta)}(\epsilon_i)$ stop at $(x_i, r_i)$ on the equilibrium curve, in which $r_i=1.5+2.5/(3x_i+1)$. Again the continuity of $\epsilon$ results in traversal for all the points on $x=x_i$, $r_c \in [r_i, r^*_0]$ by $G_{(x^*_0,r^*_0+\delta)}(\epsilon)$. Therefore, we conclude that $G_{(x^*_0,r^*_0+\delta)}(\epsilon)$ can go through any point in the region $\{(x, r_c): x\in (x^*_0, 1], r_c \in [1.5+2.5/(3x+1), r^*_0]\}$. Let $(x^*_0, r^*_0)\to (1/12, 3.5)$, we come to the conclusion that the trajectories $G_{(x^*_0,r^*_0+\delta)}(\epsilon)$ cover the whole region over the equilibrium curve, which indicates the existence of control laws to reach to any desired $(x_1, r_1)$ that is over the equilibrium curve.

(ii) In \textbf{Fig. S1B}, we prove the existence of control laws when $(x_0, r_0)$ is over the equilibrium curve while $(x_1, r_1)$ is beneath the equilibrium curve. (a) Firstly, we can control any given $(x_0, r_0)$ evolves to a stable fixed point $(x^*_0, r^*_0)$ that can arbitrarily approaches to $(1, 2.125)$ using time-dependent switching control laws with control function $f_1=(P_c-P_d)(0-(4P_c-P_d))$ and a proper relative feedback speed $\epsilon$, as already proved in (i). (b) Define $G_{(x^*_0,r^*_0-\delta)}(\epsilon)$ as the set of trajectories that begin from $(x^*_0,r^*_0-\delta)$ with a certain control function $f_2$ and relative feedback speed $\epsilon$. Let $f_2=(P_c-P_d)(2-(2P_c-P_d))$. Similarly  we can prove that $G_{(x^*_0,r^*_0-\delta)}(\epsilon)$ goes through all the points in the region $\{(x, r_c): x\in [0, x^*_0), r_c \in [1.5, r^*_0-\delta)\}$. Randomly choose a $r_i<r^*_0$, then we can reach to any point on $r=r_i$, as shown in the figure. Finally, for any final state $(x_1, r_1)$ beneath the equilibrium curve that satisfies $r_1 \ge r^*_0-\delta$, there exists a unique trajectory which belongs to the phase space of the co-evolutionary system with control function $f_3=(P_c-P_d)((2P_c-P_d)-1.2)(2-(2P_c-P_d))$ and a sufficiently large feedback speed $\epsilon$, that goes through $(x_1, r_1)$. Assume the intersection of this trajectory and $r=r_i$ is $(x_i, r_i)$, then we have a possible control path from $(x^*_0,r^*_0-\delta)$ to $(x_i, r_i)$ and finally reaches to $(x_1, r_1)$. In summary, we can always find a group of control laws given any $(x_0, r_0)$ that is over the equilibrium curve and $(x_1, r_1)$ that is beneath the equilibrium curve. 

Combining (i)(ii) which actually provide a possible control loop in phase space, we constructively prove the existence of control laws given any initial state $(x_0, r_0)$ and final state $(x_1, r_1)$, as long as the initial fraction of cooperators $x_0$ is not too small.